\documentclass[11pt,twoside]{article}  
\usepackage{adassconf}
\usepackage{asp2006}
\begin{document}   
\paperID{O7A.5}
\title{Skyalert: Real-time Astronomy for You and Your Robots}
\author{R.~D.~Williams, S.~G.~Djorgovski, A.~J.~Drake, M.~J.~Graham, A.~Mahabal}
\affil{California Institute of Technology, Pasadena, California}
\contact{Roy Williams}
\email{roy@cacr.caltech.edu}
\paindex{Williams, R.D.}
\authormark{Williams, Djorgovski, Drake, Graham, Mahabal}
\begin{abstract}          
Skyalert.org is a web application to collect and disseminate observations about time-critical astronomical transients, and to add annotations and intelligent machine-learning to those observations. 
The information is ``pushed'' to subscribers, who may be either humans (email, text message etc) or they may be 
machines that control telescopes. Subscribers can prepare precise ``trigger rules'' to decide which events should 
reach them and their robots, rules that may be based on sky position, or on the specific vocabulary of 
parameters that define a particular type of observation.
Our twin thrusts are automation of process, and discrimination of interesting events.
 \end{abstract}

An important qualitative change is happening in astronomy.  The ability to survey large areas of the sky deeply and repeatedly, and to process the data in real time, has opened a whole new field of research: the time domain astronomy.  The sky is now studied as an ever-changing collection of dynamical phenomena, with moving objects (e.g., asteroids, extrasolar planets, etc.) and objects that change in brightness (e.g., many types of variable stars, quasars, microlensing events, etc.) or appear in an explosive manner (e.g., supernovae, gamma-ray bursts, various optical transients). Many of these phenomena can be understood only through time-domain studies, and generally require a rapid follow-up using a variety of ground-based telescopes.  Thus, there is a critical need to enable a rapid dissemination of events discovered by synoptic sky surveys or various specialized experiments and space missions, to the broader astronomical community equipped to perform timely follow-up observations. We expect to handle events {\bf automatically} so that machines can work on our behalf, and we expect event delivery with personalized {\bf discrimination} of which are interesting and which are not.

Skyalert.org is a prototype system that scales up to many different event streams, and scales to large numbers of events, large numbers of event authors (streams), and large numbers of event subscribers. The new system will work well with amateur astronomers and students, as well as professional astronomers scheduling follow-up on the big NSF-funded telescopes, operated by NOAO and its partners, or on the emerging world-wide robotic telescope networks.

{\bf Skyalert Features}\\
The challenge of Skyalert is to send only the events that are wanted, and to provide within that event packet the information that makes further decisions possible with speed and accuracy.

{\it Generality}: Many new event streams are coming online or are planned for the future. Thus, it is time to deploy a system that can work with events in general, rather than a custom solution for each survey or spacecraft. Skyalert encourages adoption of a standard framework and interoperability, avoiding duplication of effort in implementing similar systems.

{\it Subscription}: From Skyalert, subscribers will get event packets immediately, through a choice of channels, content, and criterion. Subscribers can get a custom event feed, for example: all events from Catalina stream that are brighter than magnitude 16, or all Swift events that are probably a GRB. Joint trigger rules are possible ({\it ``Skymapper events which have a SDSS detection at least 2 magnitudes fainter''}. When an event satisfies a trigger, an action is taken, and these pieces of ``action code'' are easy to plug in to the Skyalert system. Actions can be: automated robotic follow-up, archive follow-up, sending a message. Actions can emit events, which can cause further actions.

{\it Annotation and Wiki-based Notices}: Each event has a web-page, a portfolio of data: a collection including spectra, cutouts, etc, from many different surveys and observations. Users can add comments and data. The evidence is thus collected, for past evidence of variability, for nearness to a galaxy, for excess infrared flux, and so on, that can help to classify that source. Trusted decisions can be made because all the information is present.

{\it Intelligence}: Just as an annotator component can get archival images for the event portfolio, so it can analyze data objects already present, for example comparing FITS images to detect motion of a source, or reduction of a spectrum to a single number of interest. These high-level criteria can be used to trigger action, as with any other parameter in the portfolio.

{\it Discovery}: We will use the international distributed Registry of the Virtual Observatory to enable publishing and discovery of event streams, so that all interested parties can subscribe to future events and browse past events from that stream. When a new event stream is created, its detailed metadata will be written and recorded by the author in a standard format.

{\it Privacy}: The Skyalert system implements virtual organizations so that some event streams can be private. A subscriber can belong to one or more of these groups, and a stream can be visible only to members of one or more groups. Furthermore, the Skyalert software is open-source and available to anyone who would like to implement it privately.

{\it Rapid Follow-up}: Real-time follow-up observations are crucial to a full understanding of the science of transients.  The Skyalert team is already collaborating with event providers (GCN, CRTS, OGLE, MOA), building new collaborations with providers of events (LOFAR, Skymapper, ATA, MWA, etc), with the amateur astronomy community (AAVSO), and with owners of robotic telescopes (LCOGT, HTN, IGO, NOAO, etc).

{\it Technology}: Skyalert is built with Django, Python, mySQL, and VOEvent. Skyalert uses modern transport such as RSS, Twitter, and XMPP Instant Message, as ways to inform humans; and with protocols such as SOAP, XML, and TCP socket for communication with machines. 

{\bf How it Works}\\
VOEvent is an international XML standard for transmitting information about a recent astronomical transient, with a view to rapid follow-up. A VOEvent packet contains one or more of the ``who, what, where, when, how, why'' of a detected event. Citations to previous VOEvents may be used to place each event in its correct context. 
In the VOEvent system, as defined by the International Virtual Observatory (IVOA), events are XML packets of a certain structure. A collection of like events is characterized as an Òevent streamÓ, meaning they all have the same project team, equipment, measured parameters, etc. An event stream might consist of gamma transients reported by the SWIFT satellite, or the transients from the OGLE microlensing survey. 
The metadata defining a stream includes the parameter names that will be used in each event, with their description, units, semantic content descriptor (UCD), and datatype. We will use this stream-specific vocabulary of keyword-parameters to build decision rules about which subscribers get which events,  also to trigger processes such as fetching archival data and putting an observation request into a telescope queue.
\begin{figure} 
\plotone{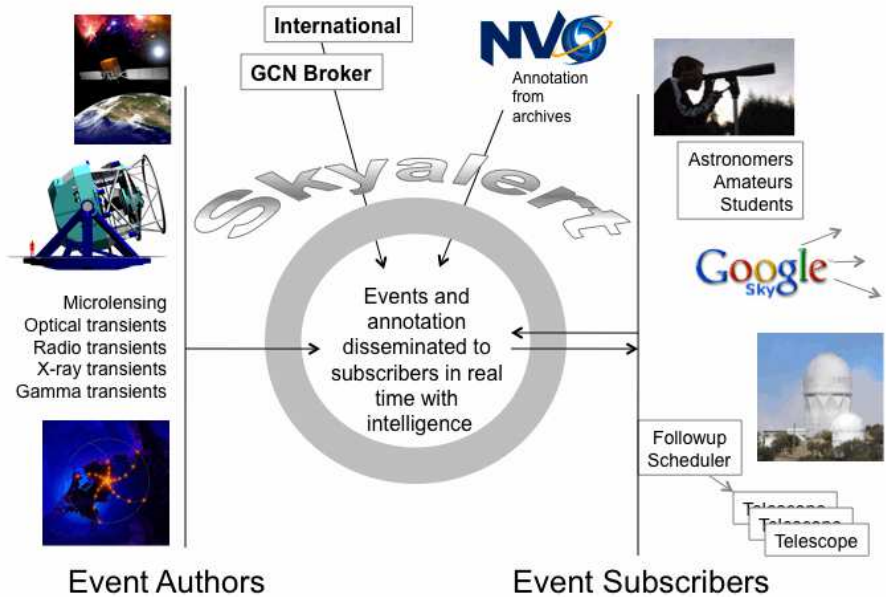} 
\vspace{-1 cm}
\caption{
The Skyalert system takes in events from authors and from other event brokers. These are compared to custom filters (ÔtriggersÕ) that subscribers have defined, and delivery will be in real time (push) via email, socket, text message, etc. Subscribers can be humans or automated systems. Events are annotated from archives and follow-up events may be sent out.} 
\label{fig:O7A.5_1} 
\end{figure} 
Skyalert makes powerful use of a particular part of the VOEvent specification: the Parameters. A parameter, in the sense of VOEvent, is a keyword-value pair, together with the metadata to give that meaning: the units, semantic meaning (via UCD), the datatype (float, int etc), and descriptive text. In our proposed work, we split the definition of the parameter from its actual value, so that a ``Parameter Template'' is defined before any events appear, and therefore the event itself need only contain the value of the parameter. Subscribers will have defined their trigger rules in terms of these parameter templates ({\it ÒI want events where $gmag < 17$Ó}), so that automated decisions and actions can be made immediately when the event actually arrives.

 When events arrive, subscribers are notified (if the event satisfies their trigger rule), but also follow-ups are initiated, that may be with a real or a virtual observatory. The real observatory slews the telescope, makes an observation, returns data as a VOEvent that cites the original one; the virtual observatory follow-up fetches archived survey data such as images, magnitudes, light curves, spectra etc, which is also returned as a VOEvent. All the collected information (original, follow-up, archival) will be presented together in VOEventNet with a rich web page that allows comments. Furthermore, subscribers will be able to build joint triggers for event subscription, that use multi-sourced data, for example {\it ÒI want Catalina transient events where the Catalina measurement is at least 2 magnitudes brighter than that from the Sloan surveyÓ.}

A key feature of Skyalert is the sophisticated specification of how events are to be received. Subscribers will register at the web site, and be able to browse and choose available event streams, then they can create one or more criteria (or Òtrigger rulesÓ) to decide which events to receive. For example an X-ray observatory called {\it Xobs} might include in each event: a geometry section, about sun and moon angles, and also flux counts in various energy bands. A subscriber could then trigger on an event by building an expression like this for a trigger rule:
\begin{eqnarray*}\tt\nonumber
Xobs["Geometry"]["Sun\_angle"] > 40\, and\, Xobs["3to8keV"]["Peak Flux"] > 3.e8 
\end{eqnarray*}
Skyalert uses these expressions in a Python sandbox to decide which events pass which triggers.

Our objective is to encourage robotic follow-up in near real time to catch the most exciting, largely unknown, physics of rapid transients. Once such observations are made, we encourage submission of that data to Skyevent as events, which will be added to the web page for the original event, and of course the follow-up event can be sent to subscribers and other event brokers. When an event becomes sufficiently ``interesting'' (i.e. passes stringent trigger rules), it will reach the attention of humans, who may comment and annotate the event web page.  When all the data is available, and colleagues have commented, there is a good basis for making fast, trusted decisions that may be expensive.

Often, unfortunately, the parameters describing the observation may not be the same as the parameters for making follow-up decisions. An observer may report multiple position measurements of the transient, to determine if it is moving, and thus an asteroid; but what would be more useful to the subscriber might be a quantity ÒasteroidnessÓ, whether this transient is 0.1 sure of being an asteroid, or 0.95 probability. Similarly with other types of classification and mining: is it a star or is it extended? What type of variable star does this light curve represent? We will work with both authors and consumers of events to build in these kinds of computational annotations, what we might call ``principal components of the decision space'' (eg asteroidness, stellarity, galactic latitude) so that subscribers can build effective discrimination triggers. 

Skyalert is currently (January 2009) a prototype, and we will make it available to subscribers and authors in the near future. Event authors and owners of robotic telescopes are especially encouraged to contact any of the authors for more information.

\acknowledgments
We are grateful to the National Science Foundation for support of this
work through DDDAS (CNS-0540369), and to NASA for support through AISRP (NNX08AD20G)


\end{document}